\documentstyle[12pt,epsfig,rotating,color,rawfonts,pictex,cite]{article}
\newcommand{\Z}{{\sf Z \!\!\! Z}}

\newcommand{\1}{{\sf 1 \!\! 1}}

\newcommand{\p}{\partial}

\newcommand{\Tr}{\mbox{Tr}}
\makeatletter
\@addtoreset{equation}{section}
\makeatother

\title{Exceptional Deconfinement in $G(2)$ Gauge Theory}

\author{M. Pepe$^a$ and U.-J. Wiese$^b$
\\ \\
$^a$ Istituto Nazionale di Fisica Nucleare and \\
Dipartimento di Fisica, Universit\`a di Milano-Bicocca, \\
3 Piazza della Scienza, 20126 Milano, Italy \\ \\
$^b$ Institute for Theoretical Physics \\
Bern University, Sidlerstrasse 5, CH-3012 Bern, Switzerland \\ \\
Dedicated to Peter Minkowski on the occasion of his $65^{th}$ birthday.}

\begin{document}

\maketitle

\vspace{-1cm}

\begin{abstract} \normalsize

The $\Z(N)$ center symmetry plays an important role in the deconfinement phase 
transition of $SU(N)$ Yang-Mills theory at finite temperature. The exceptional
group $G(2)$ is the smallest simply connected gauge group with a trivial 
center. Hence, there is
no symmetry reason why the low- and high-temperature regimes in $G(2)$ 
Yang-Mills theory should be separated by a phase transition. Still, we present
numerical evidence for the presence of a first order deconfinement phase
transition at finite temperature. Via the Higgs mechanism, $G(2)$ breaks to its
$SU(3)$ subgroup when a scalar field in the fundamental $\{7\}$ representation
acquires a vacuum expectation value $v$. Varying $v$ we investigate how the
$G(2)$ deconfinement transition is related to the one in $SU(3)$ Yang-Mills 
theory. Interestingly, the two transitions seem to be disconnected. We also
discuss a potential dynamical mechanism that may explain this behavior.

\end{abstract}

\newpage 

\section{Introduction}

Understanding the dynamical mechanism that turns confinement at low 
temperatures into deconfinement at high temperatures is an issue of central
importance in non-Abelian gauge theories. In QCD with light quarks chiral 
symmetry is spontaneously broken at low temperatures. Chiral symmetry is 
restored in a chiral phase transition at finite temperature which separates the
confined hadronic phase from the deconfined quark-gluon plasma. In QCD with
heavy quarks, on the other hand, chiral symmetry is strongly broken explicitly
and the $\Z(3)$ center of the $SU(3)$ gauge group plays an important role
\cite{Pol78,Sus79,Sve82}. In particular, in the limit of infinite quark masses,
i.e.\ for $SU(3)$ Yang-Mills theory, the center becomes an exact discrete 
global symmetry. In this limit, one finds a weak first order deconfinement 
phase transition with spontaneous breaking of the $\Z(3)$ center symmetry in 
the high-temperature deconfined phase \cite{Cel83,Kog83,Got85,Bro88,Fuk89a}. 
Indeed, the existence of an exact center symmetry necessarily implies
the existence of a finite temperature deconfinement phase transition. Quarks
with a large (but finite) mass break the center symmetry explicitly, and hence
the symmetry reason for the phase transition disappears. Indeed, when the quark
mass reaches typical QCD energy scales, the first order deconfinement phase
transition is weakened further and turns into a crossover at a critical 
endpoint \cite{Has83}.

It is interesting to investigate the role of the center symmetry for confinement and
deconfinement (for recent reviews see \cite{Hol01,Gre03,Eng04}). For this purpose, we study
gauge theories with the exceptional gauge group $G(2)$ --- the smallest simply connected
group with a trivial center \cite{Hol03,Pep05}. The Lie group $G(2)$ has 14
generators. Eight of them correspond to the 8 gluons of the $SU(3)$ subgroup. The other
six gluons transform as $\{3\}$ and $\{\overline 3\}$ of $SU(3)$ and thus carry the same
color quantum numbers as quarks and anti-quarks. Indeed, just like the quarks in QCD, the
extra gluons render the center of $G(2)$ trivial by explicitly breaking the $\Z(3)$
symmetry of the $SU(3)$ subgroup. Due to the triviality of the center, a static $G(2)$
quark in the fundamental $\{7\}$ representation can be screened by three $G(2)$ gluons. As
a consequence, the flux string connecting static $G(2)$ quarks breaks already in the pure
gauge theory by the creation of dynamical $G(2)$ gluons. Hence, the static potential
levels off at large distances and the asymptotic string tension vanishes.  Still, just
like full QCD, the theory remains confining even without an asymptotic string
tension. This follows rigorously from the behavior of the Fredenhagen-Marcu order
parameter \cite{Fre85} in the strong coupling limit of lattice gauge theory \cite{Hol03}
and is expected to hold also in the continuum limit. It should be pointed out that in
$G(2)$ Yang-Mills theory the Polyakov loop is no longer an order parameter for
deconfinement. In particular, it is non-zero (although very small) already in the confined
phase. This is another consequence of the fact that a $G(2)$ quark can be screened by at
least three $G(2)$ gluons.

Since the center of $G(2)$ is trivial, in contrast to $SU(3)$ Yang-Mills theory, there is
no symmetry reason for a deconfinement phase transition. In particular, the low- and
high-temperature regions could be connected by a smooth cross-over. However, one cannot
rule out a finite temperature phase transition. Since there is no symmetry reason for such
a transition, it is interesting to ask what dynamics may imply it. Topological objects
like instantons, merons, monopoles, or center vortices have often been invoked to address
such questions \cite{Man74,tHo77,Mac78,tHo81,Kro87a,Kro87b,Suz92,Del94,Del95,Del96,Del98,Kov98,deF99,Eng99,Kov00,deF00a,Cea00,deF00b,Lan01,Len03,Gre06,Cos06}.
While instantons, merons, or monopoles still
exist in $G(2)$ gauge theory, due to the triviality of the center, twist sectors and the
concept of center vortices do not apply here. In any case, irrespective of a particular
topological object, the deconfinement phase transition in $G(2)$ Yang-Mills theory may
result from the large mismatch between the number of degrees of freedom in the low- and
the high-temperature regimes \cite{Hol03a,Hol03b,Pep04}. In the low-temperature phase the
gluons are confined in color-singlet glueballs whose number is essentially independent of
the gauge group. In the deconfined gluon plasma, on the other hand, these degrees of
freedom are liberated and their number is determined by the number of generators of the
gauge group. In particular, for a large gauge group there is a drastic change in the
number of relevant degrees of freedom at low and high temperatures. This can cause a first
order phase transition. Indeed, while $(3+1)$-d $SU(2)$ Yang-Mills theory has a second
order deconfinement phase transition in the universality class of the 3-d Ising model,
higher $SU(N)$ groups lead to first order transitions of a strength increasing with
$N$. However, while changing the size of the gauge group, one simultaneously changes the
center $\Z(N)$ and thus the possibly available universality class. In order to disentangle
the effects of changing the size of the group and changing the center, $Sp(N)$ Yang-Mills
theories have been investigated \cite{Hol03a}. All symplectic groups $Sp(N)$ have the same
center $\Z(2)$ and thus, for symmetry reasons, a deconfinement phase transition must
occur. For $(3+1)$-d $Sp(1) = SU(2)$ Yang-Mills theory one finds the above-mentioned
second order transition in the 3-d Ising universality class.  Despite the fact that this
universality class is available for all $N$, $Sp(2)$, $Sp(3)$, and most likely all other
$Sp(N)$ Yang-Mills theories have a first order deconfinement phase transition. This shows
that the order of the phase transition does not follow from the nature of the center
(which is $\Z(2)$ for all $N$) but is a truly dynamical issue. While $Sp(2)$ (which has 10
generators) leads to a weak first order phase transition, $Sp(3)$ (with 21 generators) has
a much stronger transition. We attribute this to the larger number of liberated gluons in
the deconfined phase of $Sp(3)$ Yang-Mills theory. These arguments suggest that Yang-Mills
theory with the gauge group $G(2)$ (which has 14 generators) should also have a first
order deconfinement phase transition. Indeed, in this paper we present numerical evidence
for a first order deconfinement phase transition in $G(2)$ Yang-Mills theory. First
results on this subject have been published in \cite{Pep05}.

Since the gauge group $SU(3)$ has the nontrivial center $\Z(3)$, for symmetry reasons it
must necessarily have a deconfinement phase transition. However, in contrast to the
$SU(2)$ case, for $SU(3)$ no universality class seems to be available. In particular, the
$(4 - \varepsilon)$-expansion does not reveal a $\Z(3)$-symmetric fixed point. This led
Svetitsky and Yaffe to conjecture that the deconfinement phase transition should be first
order \cite{Sve82}. Indeed, Monte Carlo simulations of $SU(3)$ Yang-Mills theory on the
lattice show a weak first order transition. Similarly, the 3-d 3-state Potts model also
has a weak first order transition \cite{Kna79,Blo79,Her79,Wu82,Fuk89b,Gav89}. It is
interesting to ask if the first order deconfinement phase transition in $SU(3)$ Yang-Mills
theory is again due to a large number of deconfined gluons, or if it is merely an
unavoidable consequence of the nontrivial center symmetry. In this paper, we address this
question by interpolating between $SU(3)$ and $G(2)$ Yang-Mills theory. For this purpose,
we spontaneously break $G(2)$ down to $SU(3)$ with a Higgs field in the fundamental
$\{7\}$ representation. When the Higgs field picks up a vacuum value $v$, it gives mass to
the 6 additional $G(2)$ gluons, while the ordinary 8 $SU(3)$ gluons remain massless. The
mass of the additional gluons increases with $v$, such that for $v \rightarrow \infty$
they are completely removed from the spectrum.  In this limit the $G(2)$ Yang-Mills-Higgs
theory reduces to $SU(3)$ Yang-Mills theory.

\begin{figure}[htb]
\begin{center}
\epsfig{file=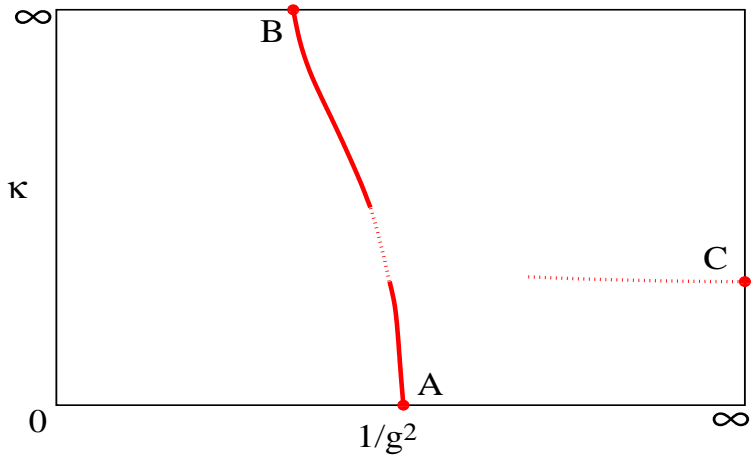,width=10.cm,height=6cm}
\caption{Phase diagram in the parameter space $(1/g^2,\kappa)$. The $\kappa=0$ and
  $\kappa=\infty$ axes correspond to $G(2)$ and $SU(3)$ Yang-Mills theories,
  respectively. The $1/g^2=\infty$ limit leads to the $SO(7)$ spin model. The dotted lines
  correspond to crossover regions or, eventually, to very weak phase transitions.}
\label{phasediag}
\end{center}
\end{figure}

The numerical simulations to be presented below result in the phase diagram of figure
\ref{phasediag}. The two axes correspond to the inverse gauge coupling $1/g^2$ and the
hopping parameter $\kappa$ of the fixed length Higgs field in the $\{7\}$
representation. The simulations are performed at fixed Euclidean time extent 
$N_t = 6$. Hence, varying $1/g^2$ effectively changes the physical temperature.  At 
$\kappa =\infty$ the $G(2)$ Yang-Mills theory reduces to $SU(3)$ with its weak first order
deconfinement phase transition. As $\kappa$ is lowered, in addition to the 8 $SU(3)$
gluons, 6 $G(2)$ gluons of decreasing mass begin to participate in the dynamics. Just like
dynamical quarks and anti-quarks, they transform in the $\{3\}$ and $\{\overline{3}\}$
representation of $SU(3)$ and thus explicitly break the $\Z(3)$ center symmetry. As in
full QCD, this leads to a weakening of the deconfinement phase transition, which even
seems to disappear at a critical endpoint. Due to finite size effects, the location of the
endpoint and even its existence cannot be determined unambiguously from our Monte Carlo
data.

At $\kappa = 0$ we are in the pure $G(2)$ gauge theory. Although $G(2)$ has a trivial
center, we find a clear first order phase transition. Despite the fact that the Polyakov
loop is not a true order parameter, in the sense that it vanishes in the confined phase, it
jumps from a very small non-zero value at low temperatures to a large value at high
temperatures.  Hence, although the low- and high-temperature regions are not distinguished
by different symmetry properties, the transition shares the important features of a
deconfinement phase transition. Of course, as we stressed before, in $G(2)$ the
low-temperature confined phase has string-breaking and thus no asymptotic string
tension. We attribute the existence of the transition to the large mismatch between the
number of the relevant degrees of freedom in the two phases.  While there is a small number of
glueball states at low temperatures, there are 14 deconfined $G(2)$ gluons at high
temperatures. As $\kappa$ increases, at some point 6 $G(2)$ gluons pick up a mass and are
progressively removed from the dynamics. This reduces the mismatch in the number of
degrees of freedom and implies a weakening of the $G(2)$ deconfinement phase transition.
Again, due to finite size effects, the existence of a critical endpoint cannot be derived
unambiguously from our Monte Carlo data. The data suggest the existence of an endpoint and
thus of a crossover region (indicated in figure 1 by the dotted part of the line between
the points A and B), but we cannot rule out a very weak first order transition. In any
case, the $G(2)$ transition disappears (or weakens substantially) before the $SU(3)$
transition appears. This suggests that, unlike the $G(2)$ transition, the $SU(3)$
transition is not due to a large mismatch in the number of relevant degrees of
freedom. For $SU(3)$, due to the nontrivial $\Z(3)$ center symmetry, a deconfinement
phase transition must exist. It is of first order because no universality class (which should
be visible in the $(4 - \varepsilon)$-expansion) seems to be available.

At $1/g^2 = \infty$ the theory reduces to an $SO(7)$-invariant nonlinear $\sigma$-model
which shows spontaneous symmetry breaking down to $SO(6)$ at the critical point
denoted by C in figure 1. The dotted line emerging from this point is a 
crossover or perhaps a phase transition which we have not investigated in
much detail. Despite the fact that $G(2)$ has a trivial center, in the 
deconfined phase there are metastable minima of the effective potential for the
Polyakov loop. Although these minima are irrelevant in the infinite volume 
limit, they may influence the physics at finite volume. In order to correctly 
interpret the results of numerical simulations at high temperatures, the 
existence of these metastable minima must be taken into account. In particular,
they obscure the physics around the dotted line connected to C in figure \ref{phasediag}.

The rest of the paper is organized as follows. In section 2 the basic features
of the group $G(2)$ are discussed and $G(2)$ gauge theories are introduced both
in the continuum and on the lattice. Deconfinement in $G(2)$ Yang-Mills theory
is investigated and Monte Carlo results are presented in section 3. In section 
4 the 1-loop effective potential for the $G(2)$ Polyakov loop is derived
analytically at high temperatures. In section 5 a Higgs field is added in order
to break $G(2)$ down to $SU(3)$ and thus to investigate how the exceptional 
deconfinement of $G(2)$ is related to the familiar deconfinement of $SU(3)$ 
Yang-Mills theory. Finally, section 6 contains our conclusions. The Monte Carlo
algorithm for updating $G(2)$ lattice gauge theory is described in an appendix.

\section{$G(2)$ Gauge Theory}

In this section we discuss some basic properties of the group $G(2)$, and
construct $G(2)$ Yang-Mills and gauge-Higgs theories both in the continuum and
on the lattice.

\subsection{The Exceptional Group $G(2)$}

$G(2)$ is the smallest among the exceptional Lie groups $G(2)$, $F(4)$, $E(6)$,
$E(7)$, and $E(8)$. It is physically interesting because it has a trivial 
center and contains the group $SU(3)$ as a subgroup. The group 
$G(2)$ has rank 2, 14 generators, and the fundamental representation is 
7-dimensional. Since $G(2)$ is real it must be a subgroup of the rank 3 group 
$SO(7)$ which has 21 generators. The $7 \times 7$ real orthogonal matrices 
$\Omega$ of $SO(7)$ obey the constraint
\begin{equation}
\Omega_{ab} \Omega_{ac} = \delta_{bc}
\end{equation}
and have determinant 1. The elements of $G(2)$ satisfy the additional cubic 
constraint
\begin{equation}
\label{cubic}
T_{abc} = T_{def} \Omega_{da} \Omega_{eb} \Omega_{fc},
\end{equation}
where $T$ is a totally anti-symmetric tensor whose non-zero elements follow by 
anti-symmetrization from
\begin{equation}
\label{tensor}
T_{127} = T_{154} = T_{163} = T_{235} = T_{264} = T_{374} = T_{576} = 1.
\end{equation}
Restricting to the $SU(3)$ subgroup, the fundamental and adjoint $G(2)$ 
representations are reducible and decompose as 
\begin{equation}
\{7\} \longrightarrow \{3\} \oplus \{\overline 3\} \oplus \{1\}, \;\;\;\;\;\;
\{14\} \longrightarrow \{8\} \oplus \{3\} \oplus \{\overline 3\}.
\end{equation}
Under the subgroup $SU(3)$, 8 of the 14 $G(2)$ gluons transform like ordinary 
gluons, i.e.\ as an $\{8\}$ of $SU(3)$. The remaining 6 $G(2)$ gauge bosons 
break up into $\{3\}$ and $\{\overline 3 \}$ and thus have the color quantum 
numbers of ordinary quarks and anti-quarks. Although these objects are vector 
bosons, they have similar effects as quarks in full QCD. In particular, they 
explicitly break the $\Z(3)$ center symmetry and make the center of $G(2)$
trivial. Due to the trivial center, the decomposition of the $G(2)$ tensor 
product of three adjoint representations contains the fundamental 
representation
\begin{eqnarray}
\{14\} \otimes \{14\} \otimes \{14\}&=&\{1\} \oplus \{7\} \oplus 5 \; \{14\} 
\oplus 3 \; \{27\} \oplus 2 \; \{64\} \oplus 4 \; \{77\} \oplus 3 \; \{77'\}
\nonumber \\
&\oplus&\{182\} \oplus 3 \; \{189\} \oplus \{273\} \oplus 2 \; \{448\}.
\end{eqnarray}
As a consequence, three $G(2)$ gluons can screen a single $G(2)$ quark, and
hence the flux string can break already in the pure gauge theory. Further 
properties of the group $G(2)$ can be found in \cite{Beh62,Gun73,McKay,Mac01,Hol03}.

\subsection{$G(2)$ Gauge Theories in the Continuum}

The Lagrangian for $G(2)$ Yang-Mills theory takes the standard form
\begin{equation}
{\cal L}_{YM}[A] = \frac{1}{2 g^2} \mbox{Tr} F_{\mu\nu} F_{\mu\nu},
\end{equation}
where the field strength
\begin{equation}
F_{\mu\nu} = \p_\mu A_\nu - \p_\nu A_\mu + [A_\mu,A_\nu],
\end{equation}
is derived from the vector potential
\begin{equation}
A_\mu(x) = i g A_\mu^a(x) \frac{\Lambda_a}{2}.
\end{equation}
Here $\Lambda_a$ are the 14 generators of $G(2)$ with normalization 
$\Tr\, [\Lambda_a \Lambda_b] = 2\delta_{ab}$. The Lagrangian is invariant 
under non-Abelian gauge transformations
\begin{equation}
A_\mu' = \Omega (A_\mu + \p_\mu) \Omega^\dagger, \;\;\;\; \Omega(x) \in G(2). 
\end{equation}

Let us also add a Higgs field in the fundamental $\{7\}$ representation in 
order to break $G(2)$ spontaneously down to $SU(3)$. Then 6 of the 14 $G(2)$
gluons pick up a mass proportional to the vacuum value $v$ of the Higgs
field, while the remaining 8 $SU(3)$ gluons are unaffected by the Higgs 
mechanism and are confined inside glueballs. For large $v$ the theory reduces 
to $SU(3)$ Yang-Mills theory. Hence, by varying $v$ one can interpolate between
$G(2)$ and $SU(3)$ Yang-Mills theory. The Lagrangian of the $G(2)$ gauge-Higgs 
model is given by
\begin{equation}
{\cal L}_{GH}[A,\varphi] = {\cal L}_{YM}[A] + 
\frac{1}{2} D_\mu \varphi D_\mu \varphi + V(\varphi),
\end{equation}
where $\varphi(x) = (\varphi^1(x),\varphi^2(x),...,\varphi^7(x))$ is the 
real-valued Higgs field,
\begin{equation}
D_\mu \varphi = (\p_\mu + A_\mu) \varphi,
\end{equation}
is the covariant derivative and
\begin{equation}
V(\varphi) = \lambda (\varphi^2 - v^2)^2
\end{equation}
is the scalar potential.

The ungauged Higgs model with the Lagrangian
\begin{equation}
{\cal L}_{H}[\varphi] = \frac{1}{2} \p_\mu \varphi \p_\mu \varphi + V(\varphi).
\end{equation}
even has an enlarged global symmetry $SO(7)$ which is spontaneously broken to 
$SO(6)$ in a second order phase transition. There are $21 - 15 = 6$ massless 
Goldstone bosons. Gauging only the $G(2)$ subgroup of $SO(7)$ we break the 
global $SO(7)$ symmetry explicitly, and the global $SO(6) \simeq SU(4)$ 
symmetry turns into a local $SU(3)$ symmetry. Hence, a Higgs in the $\{7\}$ 
representation of $G(2)$ breaks the gauge symmetry down to $SU(3)$. The 6 
massless Goldstone bosons are eaten and become the longitudinal components of
$G(2)$ gluons which pick up a mass $M_G = g v$, while the remaining 8 gluons 
are those familiar from $SU(3)$ Yang-Mills theory.

\subsection{$G(2)$ Gauge Theories on the Lattice}

The construction of $G(2)$ Yang-Mills theory on the lattice is straightforward.
The link matrices $U_{x,\mu} \in G(2)$ are group elements in the fundamental 
$\{7\}$ representation. We will use the standard Wilson plaquette action 
\begin{equation}
S_{YM}[U] = - \frac{1}{g^2} \sum_\Box \mbox{Tr} \ U_\Box =
- \frac{1}{g^2} \sum_{x,\mu < \nu} \mbox{Tr} \
U_{x,\mu} U_{x+\hat\mu,\nu} U^\dagger_{x+\hat\nu,\mu} U^\dagger_{x,\nu},
\end{equation}
where $g$ is the bare gauge coupling and $\hat \mu$ is the unit-vector in the
$\mu$-direction. The partition function then takes the form
\begin{equation}
Z = \int {\cal D}U \exp(- S[U]),
\end{equation}
where the measure
\begin{equation}
\int {\cal D}U = \prod_{x,\mu} \int_{G(2)} dU_{x,\mu},
\end{equation}
is a product of local $G(2)$ Haar measures for each link. Both the action and 
the measure are invariant under gauge transformations
\begin{equation}
U'_{x,\mu} = \Omega_x U_{x,\mu} \Omega^\dagger_{x+\hat\mu},
\end{equation}
with $\Omega_x \in G(2)$. Despite the fact that the $G(2)$ flux string can
break and the asymptotic string tension vanishes, it has been shown that $G(2)$
Yang-Mills theory is still confining, at least in the strong coupling limit 
\cite{Hol03}. The same is true in full QCD.

Let us now add the scalar Higgs field in the fundamental $\{7\}$ representation
of $G(2)$ to the lattice theory. For this purpose, we introduce a real-valued
7-component vector $\varphi_x$ at each lattice point $x$ and we consider the 
scalar potential $V(\varphi_x) = \lambda (\varphi_x^2 - v^2)^2$. It is 
convenient to take the limit $\lambda \rightarrow \infty$ and (after rescaling)
to choose $v = 1$ (in lattice units). Hence, the scalar field is then described
by a 7-component unit-vector. Adding a kinetic hopping term for the scalar 
field, the gauge-Higgs action now reads
\begin{equation}
S_{GH}[U,\varphi] = S_{YM}[U] - \kappa \sum_{x,\mu} 
\varphi^T_x U_{x,\mu} \varphi_{x + \hat \mu},
\end{equation}
where $\kappa$ is the hopping parameter. Again, the action is gauge invariant 
since
\begin{equation}
\varphi'_x = \Omega_x \varphi_x.
\end{equation}

\section{Deconfinement in $G(2)$ Yang-Mills Theory}

In this section we introduce the $G(2)$ Polyakov loop and point out that it is no longer
an order parameter for deconfinement in the strict sense. We then present results of Monte
Carlo simulations showing a first order deconfinement phase transition at finite
temperature.

\subsection{The Polyakov loop}

The Polyakov loop is an interesting physical quantity that distinguishes confined
from deconfined phases. In the continuum it is given by
\begin{equation}
\Phi(\vec x) = \mbox{Tr} {\cal P} \exp\left(\int_0^\beta dt \ A_4(\vec x,t)
\right).
\end{equation}
Here ${\cal P}$ denotes path ordering and $\beta = 1/T$ is the inverse 
temperature. On the lattice the Polyakov loop
\begin{equation}
\Phi_{\vec x} = \mbox{Tr} \prod_t U_{\vec x,t,4}
\end{equation}
is the trace of a path ordered product of link variables along the periodic
Euclidean time direction. Its expectation value
\begin{equation}
\langle \Phi \rangle = \frac{1}{Z} \int {\cal D}U \ \Phi_{\vec x} \exp(- S[U]),
\end{equation}
is related to the free energy $F$ of an external static quark by
\begin{equation}
\langle \Phi \rangle = \exp(- \beta F).
\end{equation}
Here $\beta = N_t$ is given by the extent $N_t$ of the Euclidean time direction. In
contrast to Yang-Mills theories with a nontrivial center, in $G(2)$ Yang-Mills theory
$\langle \Phi \rangle \neq 0$ even in the low-temperature confined phase. Hence, the free
energy $F$ of a static quark is always finite. This is a consequence of string-breaking: a
single quark can be screened by at least three $G(2)$ gluons. Since it does not vanish in
the confined phase, the Polyakov loop is no longer an order parameter for deconfinement in
$G(2)$ Yang-Mills theory. In particular, again in contrast to Yang-Mills theories with a
nontrivial center but just like in full QCD, there is no symmetry reason for a
deconfinement phase transition.  The lack of an order parameter in $G(2)$ Yang-Mills
theory is due to the fact that the low-temperature confined and high-temperature
deconfined regions are analytically connected. In particular, a priori it is possible that
there is just a cross-over and no deconfinement phase transition at all. Still, there
might be a first order phase transition unrelated to symmetry breaking. To decide if there
is a cross-over or a first order phase transition is a subtle dynamical question which
requires non-perturbative insight. Indeed, the numerical simulations to be presented below
show a first order deconfinement phase transition at finite temperature. It should be
noted that the Polyakov loop can still be used to locate the phase transition. At the
transition temperature, it jumps from a very small (but non-zero) value in the confined
phase to a large value in the deconfined phase. Recent investigations have been devoted to
the construction of effective models for the deconfining phase transition in terms of the
Polyakov loop \cite{Dum03,Pis06}.

\subsection{Results of Monte Carlo Simulations}

\begin{figure}[htb]
\begin{center}
\vskip-.2mm \hskip.4cm  (a) \hskip7.cm (b)\\
\vskip-.5cm \hskip-7.5cm
\epsfig{file=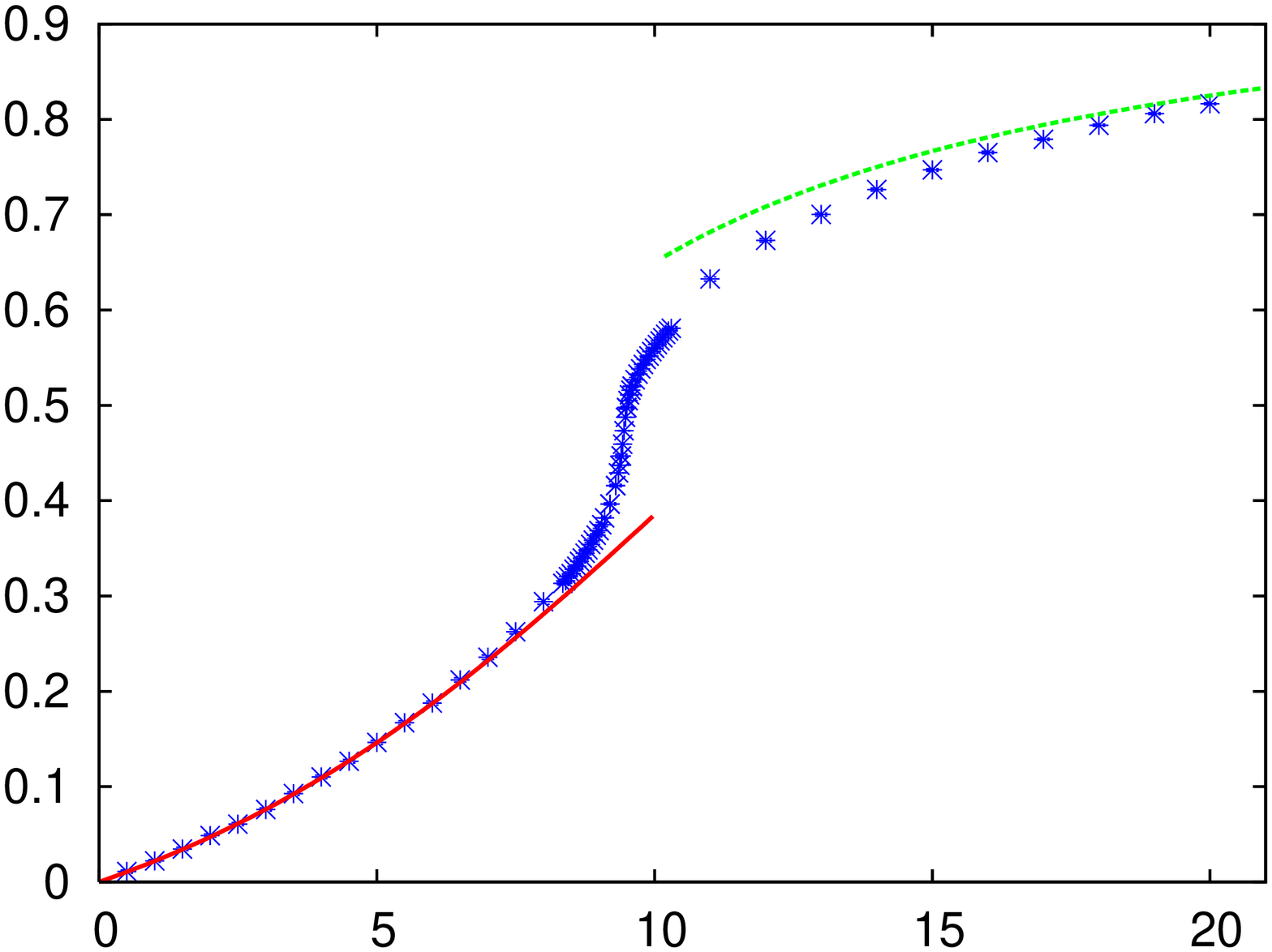,width=5.2cm,height=7.5cm,angle=-90}
\vskip -5.25cm\hskip7.7cm
\epsfig{file=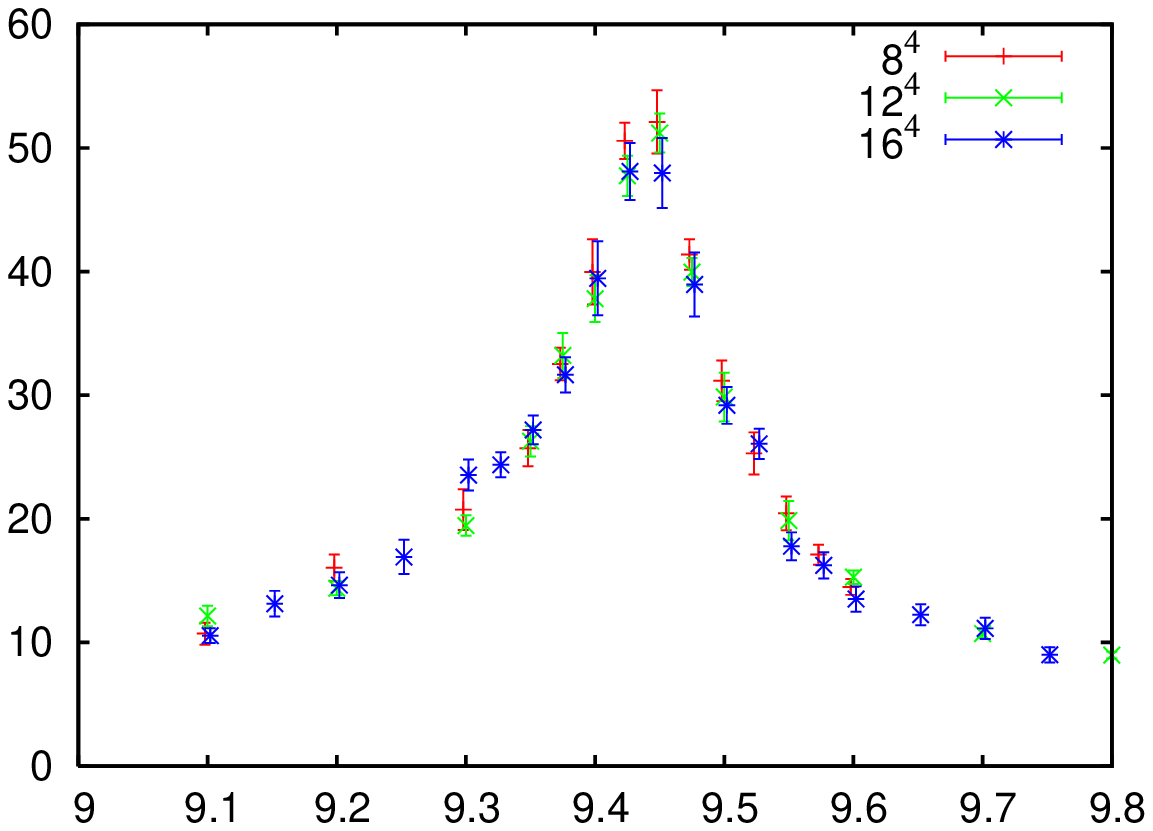,width=7.3cm,height=5.2cm}
\vskip -4.1cm
\hskip-7cm 
\begin{sideways}
{\small $\mbox{Tr} \ U_\Box/7$}
\end{sideways}
\hskip7cm {\small $C_s$}
\vskip 2.5cm
\hskip.7cm {\small $7/g^2$}
\hskip6.8cm {\small $7/g^2$}
\caption{(a): Average plaquette $\mbox{Tr} \ U_\Box/7$ as a function of the gauge coupling
  $7/g^2$. The lattice size is $8^4$. The solid and dotted lines correspond,
  respectively, to the strong and weak coupling expansions. (b): Specific heat $C_s$ in
  the crossover region.}\label{bulk}
\end{center}
\end{figure}

We have performed numerical simulations of $(3+1)$-d $G(2)$ Yang-Mills theory
on the lattice with the standard Wilson action using the algorithm described in
appendix~A. First, we have investigated the possible presence of a bulk phase
transition in the bare coupling. Figure \ref{bulk}a shows the average plaquette as
a function of the bare coupling $7/g^2$. The solid and dotted lines correspond,
respectively, to the strong coupling expansion $h(7/g^2)$ and the weak coupling expansion
$c(7/g^2)$ 
\begin{equation}
h(\frac{7}{g^2})=
\frac{1}{7} \left(\frac{1}{g^2}+\frac{1}{2 (g^2)^2}
+\frac{1}{6 (g^2)^3}-\frac{2305}{57624 (g^2)^5}\right) ,
\;\;\;\;\;
c(\frac{7}{g^2})=1-\frac{1}{2}g^2.
\end{equation}
In figure \ref{bulk}b we plot the specific heat
\begin{equation}
C_s= \frac{1}{6 N_s^3 N_t}\left( \langle S_{YM}^2\rangle -
\langle S_{YM} \rangle ^2 \right)
\end{equation}
in the region between the strong and the weak coupling regimes. The specific heat does not
increase with the volume and, as already pointed out in \cite{Pep05}, we do not find any
indication for a bulk phase transition. There is only a rapid crossover between the strong
and weak coupling regimes around $7/g^2 \approx 9.440(15)$. A recent study \cite{Cos06}
claims that the strong and the weak coupling regimes are separated by a bulk phase
transition. In our further numerical simulations we will stay on the weak coupling side of
the crossover. 
\vskip-5mm

\begin{figure}[htb]
\begin{center}
\epsfig{file=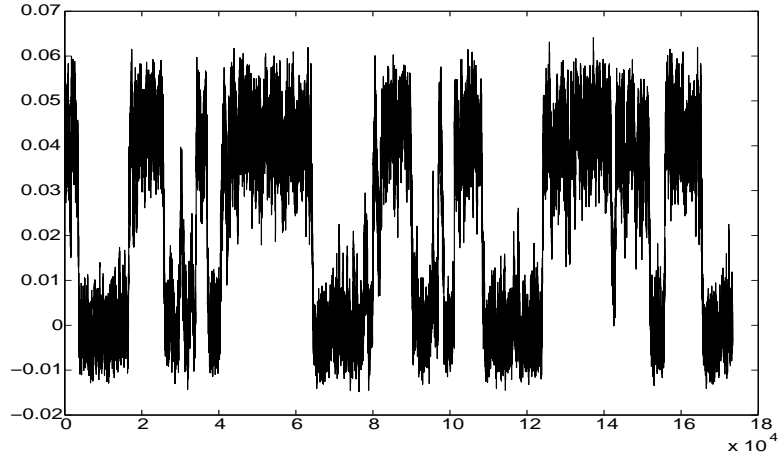,width=10cm,height=6cm}
\caption{Monte Carlo history of the Polyakov loop from a numerical simulation on a
  $20^3\times 6$ lattice at $7/g^2=9.765$.}\label{YMpoly}
\end{center}
\end{figure}

In particular, we have performed simulations at finite
temperature keeping the extent of the Euclidean time direction fixed at
$N_t = 6$. Varying the spatial lattice size $N_s$ between 12 and 20, we have
found a first order deconfinement phase transition at $7/g_c^2 \approx 9.765$. 
Figure \ref{YMpoly} shows the Monte Carlo history of the Polyakov loop which 
displays numerous tunneling events indicating coexistence of the low- and 
high-temperature phases. 

\begin{figure}[htb]
\begin{center}
\epsfig{file=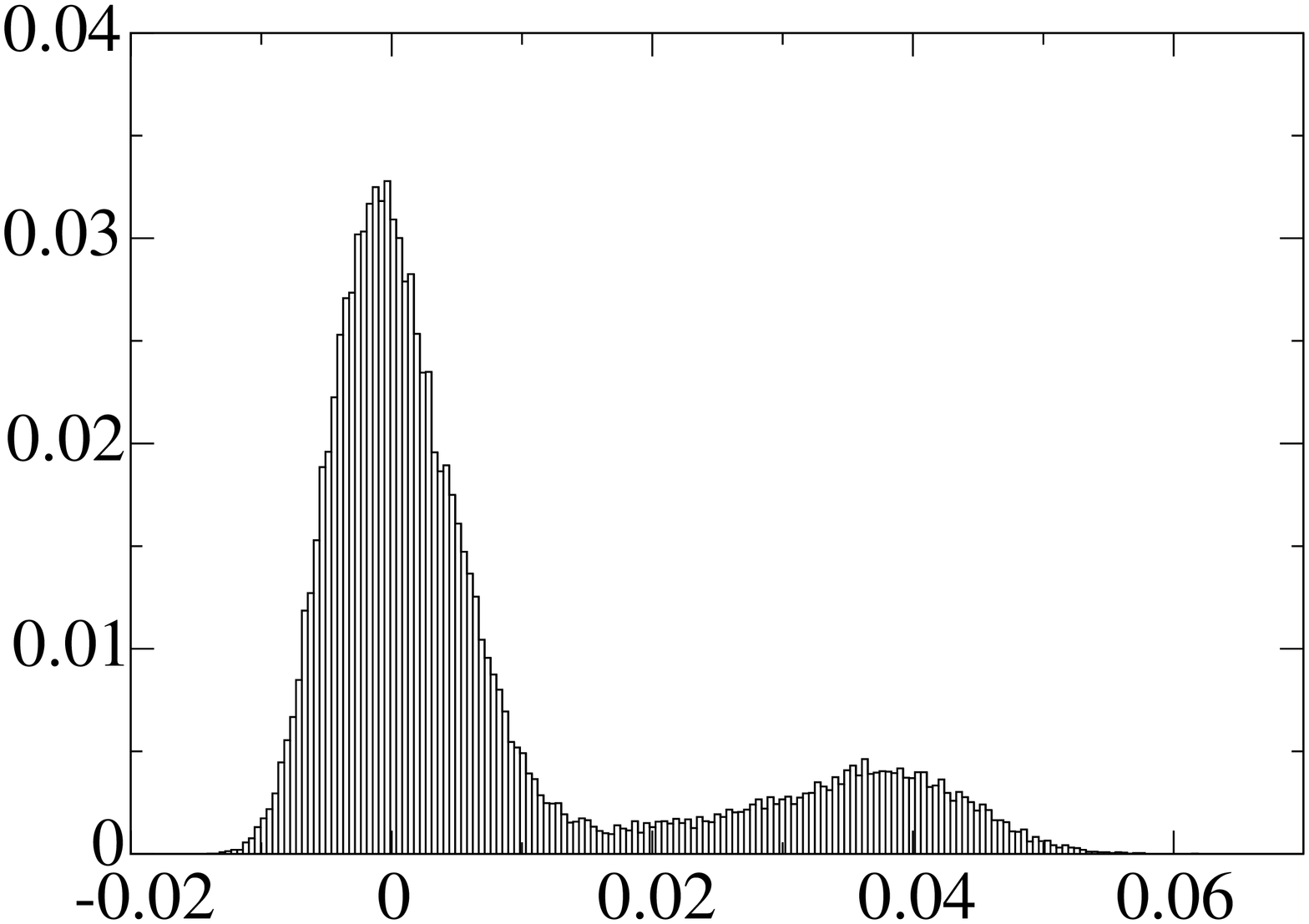,width=3.5cm,angle=-90}
\epsfig{file=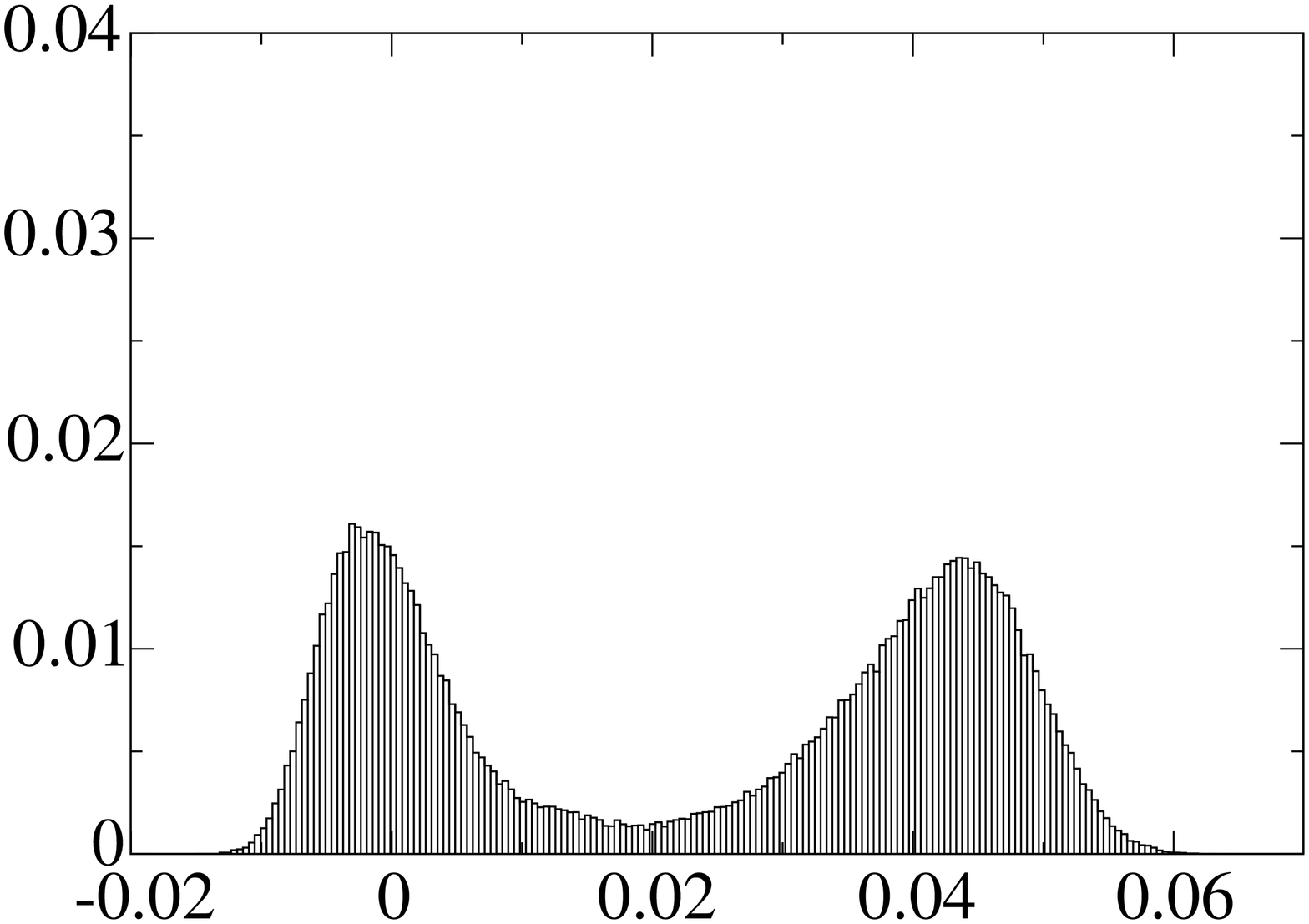,width=3.5cm,angle=-90}
\epsfig{file=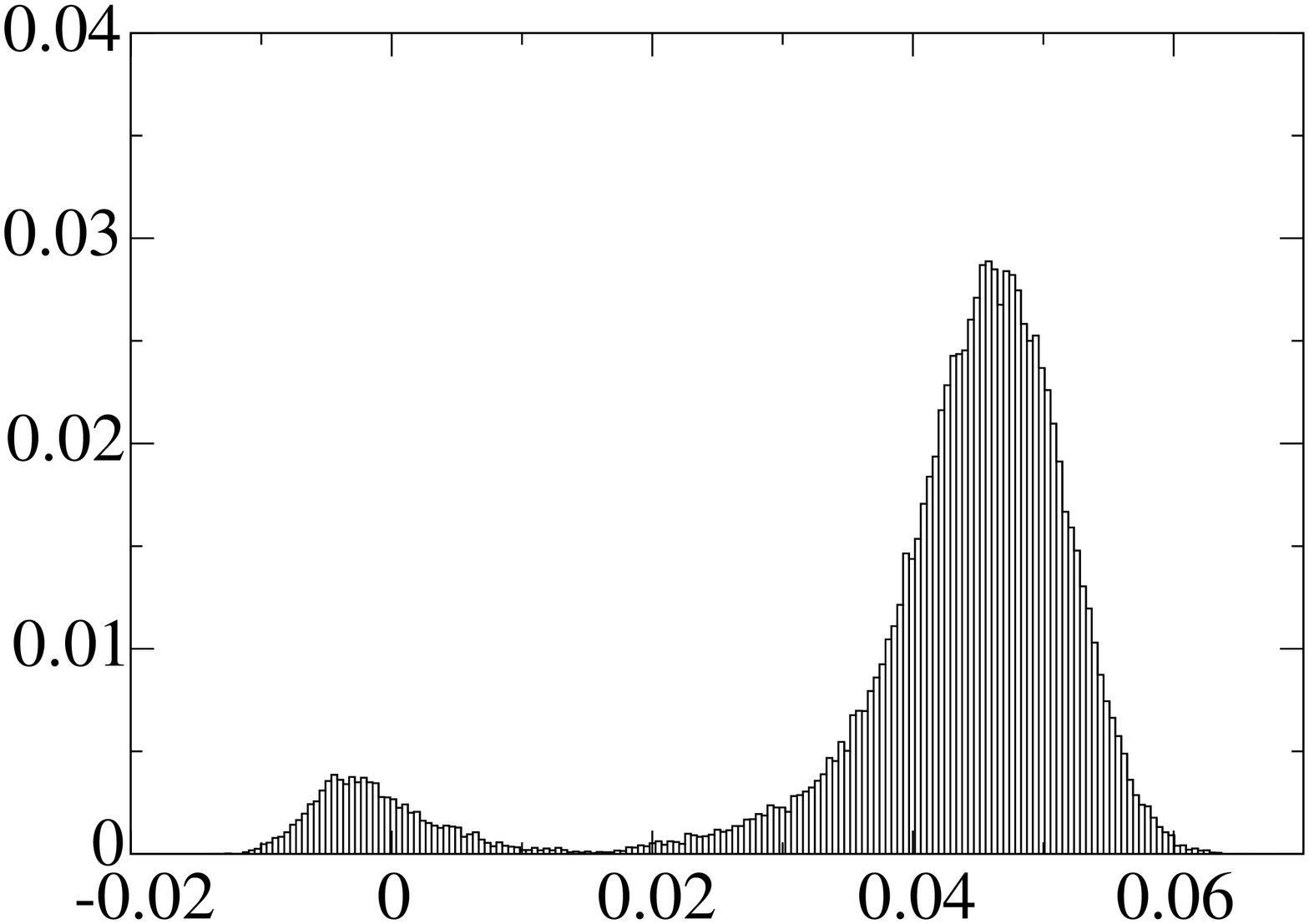,width=3.5cm,angle=-90}
\caption{Polyakov loop probability distributions in the region of the
  deconfinement phase transition in (3+1)-d $G(2)$ Yang-Mills theory. The
  temperature increases from left to right. The simulations have been performed on a
  $20^3\times 6$ lattice at the three gauge couplings $7/g^2= 9.75$, 9.765, and
  9.775 (left to right).}\label{YMhist}
\end{center}
\end{figure}

In addition, figure \ref{YMhist} shows histograms of
the Polyakov loop distribution around the deconfinement phase transition. At
low temperatures a single peak is located very close to zero, i.e.\ the free
energy of a static quark is very large (although not infinite). As one 
approaches the phase transition, a second peak emerges. This peak corresponds 
to the high-temperature phase and has a much larger value of the Polyakov loop,
i.e.\ a static quark now has a much smaller free energy. As we further increase
the temperature (by increasing $7/g^2$) the peak corresponding to the 
low-temperature phase disappears and we are left with the deconfined peak only.
We have varied $N_t$ to check that the critical bare coupling $7/g_c^2$ varies
accordingly, but we have not attempted to extract the value of the critical 
temperature in the continuum limit. 

In the high-temperature phase we have observed tunneling events between 
different minima of the effective potential for the Polyakov loop. In $SU(3)$
gauge theory these would simply represent the three different $\Z(3)$ copies of
the deconfined phase. Since the center symmetry is explicitly broken in $G(2)$
Yang-Mills theory, one would not necessarily expect this phenomenon. Indeed, 
the additional minima are not degenerate with the lowest one and hence
represent metastable phases which disappear in the infinite volume limit.
However, in a finite volume they are present and may obscure the physics if one
is not aware of them. In order to gain a better understanding of the metastable
phases, the next section contains an analytic 1-loop calculation of the 
effective potential for the Polyakov loop at very high temperatures.

\section{High-Temperature Effective Potential for the Polyakov Loop}

In this section we analytically calculate the 1-loop effective potential
for the Polyakov loop in the high-temperature limit, which turns out to have
metastable minima besides the absolute minimum representing the stable 
deconfined phase. Our calculation is the $G(2)$ analog of the calculation of
the effective potential for the Polyakov loop in $SU(N)$ Yang-Mills theory 
\cite{Wei81,Wei82,Bel90}.

In contrast to calculations at zero temperature, at finite temperature --- due 
to the compact temporal direction --- the time-component $A_4$ of the gauge 
field cannot be gauged to zero but only to a constant. Hence, perturbative 
expansions around background gauge fields with different static temporal 
components are physically different at finite temperature.

\begin{figure}[htb]
\begin{center}
\vskip-.6cm \hskip.5cm  (a) \hskip7.cm (b)\\ 
\epsfig{file=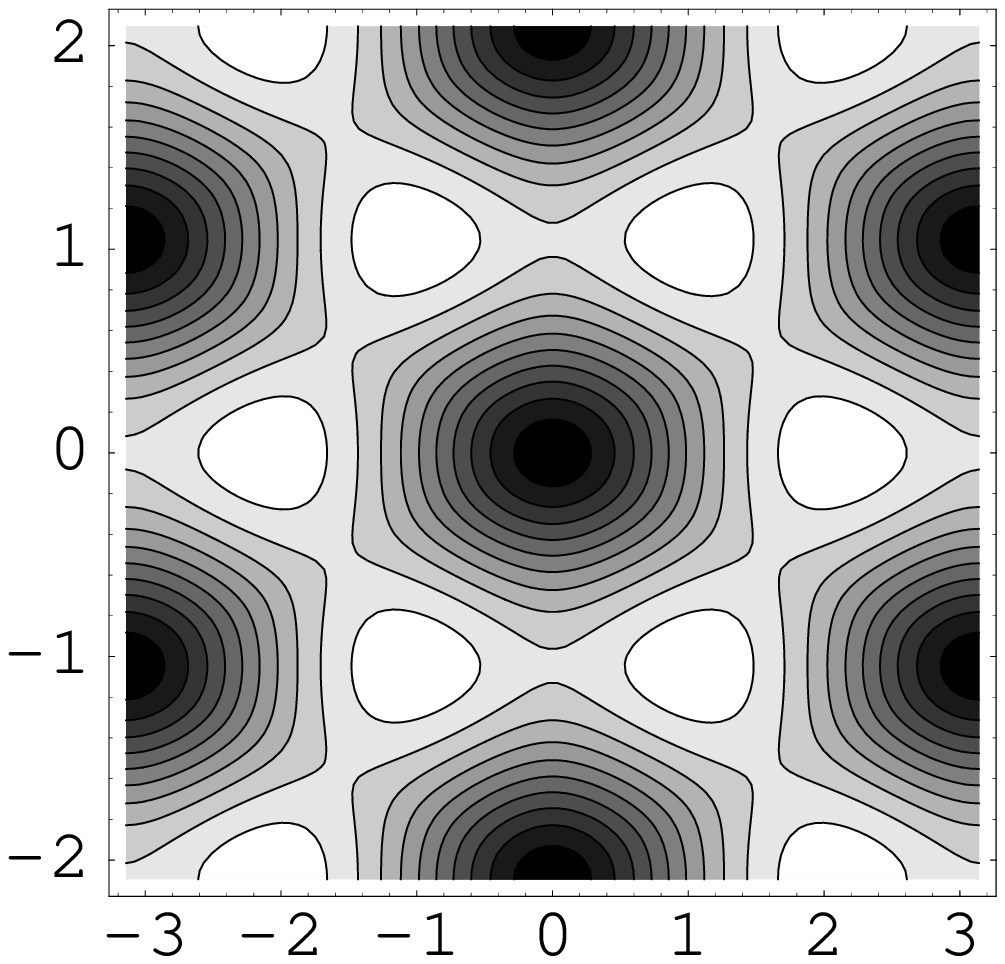,width=4.6cm}\hskip3cm
\epsfig{file=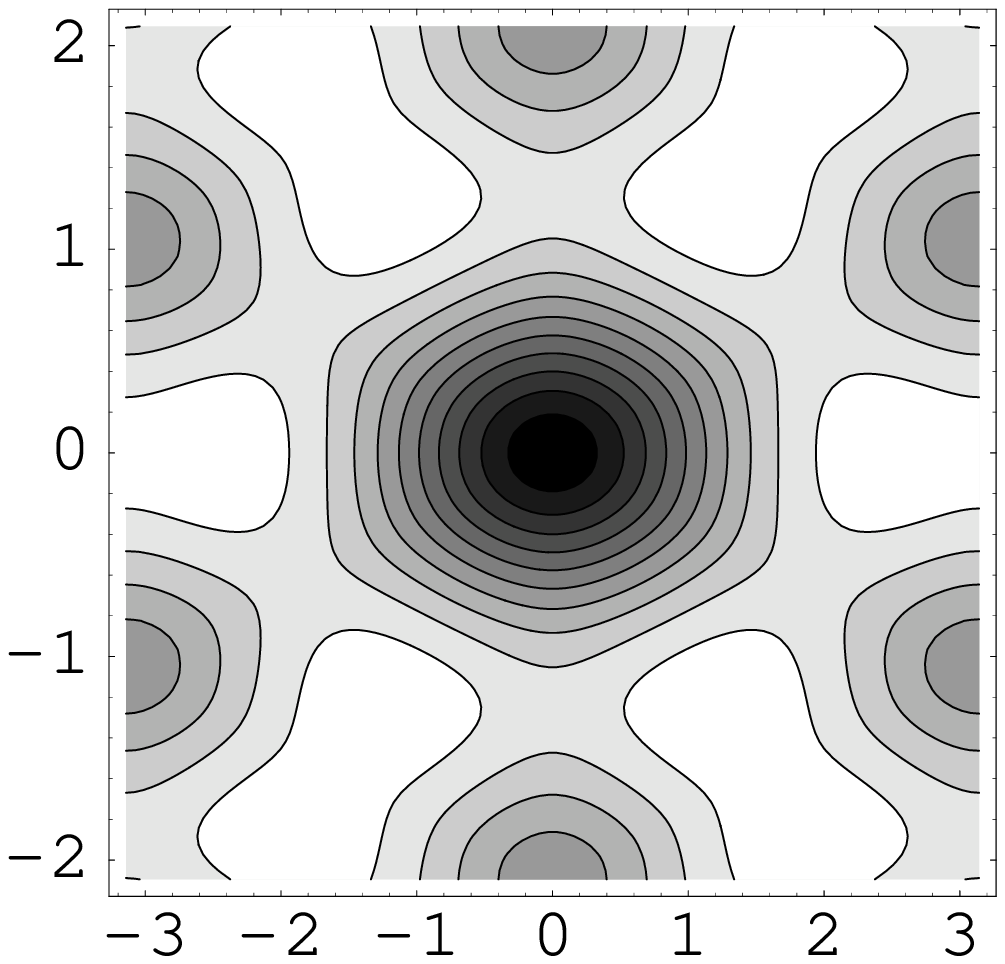,width=4.5cm}
\vskip -2.9cm
\hskip-5cm {\small $\tilde\theta_2$}\hskip7.4cm {\small $\tilde\theta_2$}
\vskip 2.1cm
\hskip.65cm {\small $\tilde\theta_1$}\hskip7.3cm {\small $\tilde\theta_1$}
\caption{Contour plots of the one-loop high-temperature effective potential for
the Polyakov loop in (a) $SU(3)$ and (b) $G(2)$ Yang-Mills theory.}
\label{effpotG2SU3}
\end{center}
\end{figure}

In order to compute the one-loop free energy in the continuum, we perturb 
around a general constant background gauge field $A_\mu^B $ given by
\begin{equation}
A_4^B  = \sqrt{2}\; \theta_1 \Lambda_3 + \sqrt{6}\;\theta_2 \Lambda_8, 
\;\;\;\;\;\;
A_i^B  = 0, 
\end{equation}
where $\theta_1$ and $\theta_2$ are the two phases characterizing an Abelian 
$G(2)$ matrix. The effective potential for the background field $A_\mu^B$ 
directly yields the effective potential for the Polyakov loop which is simply 
the time-ordered integral of $A_4^B$ along the temporal direction. We now 
fix to the covariant background gauge $D_\mu^B A_\mu = 0$, where $D_\mu^B$ is
the covariant derivative in the background field. After integrating in the 
ghost field $\chi$, the continuum Lagrangian takes the form
\begin{equation}
{\cal {L}} = \frac{1}{2 g^2} \Tr[F_{\mu\nu} F_{\mu\nu}] +
\frac{1}{\alpha} \Tr[(D_\mu^B A_\mu)^2] +
2\, \Tr[\overline{\chi} D_\mu^B D_\mu \chi],
\end{equation}
where $\alpha$ is a gauge fixing parameter. We now decompose the gauge field 
into the background field and quantum fluctuations $A_\mu^q $, i.e.
\begin{equation}
A_\mu  = A_\mu^B  + A_\mu^q.
\end{equation}
We expand the Lagrangian around the background field and keep only those terms
that are at most quadratic in $A_\mu^q $. The partition function can then be 
computed and the resulting free energy $W$ is given by
\begin{equation}
W[\theta_1,\theta_2] = \frac{1}{2} \Tr \left\{ \log \left[  
\left( \delta_{\mu\nu} (-D^B)^2 \right)+(1-\frac{1}{\alpha}) D_\mu^B D_\nu^B
\right] \right\} 
-\Tr \left[ \log(-(D^B)^2)\right].
\end{equation}
Using some relations from \cite{Bel90}, the one-loop free energy can be 
explicitly calculated and the result is
\begin{equation}\label{effpot}
W[\theta_1,\theta_2] = \frac{4\pi^2}{3 \beta^4} 
\left[
-\frac{1}{30} +\sum_{i=1}^6 B_4\left( 
\frac{C_i(\tilde\theta_1,\tilde\theta_2)}{2\pi}\right)
\right],
\end{equation}
where $B_4(x)=-1/30 +x^2 (x-1)^2$ is the fourth Bernoulli polynomial with the 
argument defined modulo 1, and $\tilde\theta_1=g \theta_1 \beta$, 
$\tilde\theta_2=g \theta_2 \beta$. The functions $C_i$ are given by
\begin{equation}
\begin{array}{lll}
C_1(\tilde\theta_1,\tilde\theta_2)= 2 \tilde\theta_1, \;\;\;\; &
C_2(\tilde\theta_1,\tilde\theta_2)=\tilde\theta_1 + 3 \tilde\theta_2, \;\;\;\; 
&
C_3(\tilde\theta_1,\tilde\theta_2)=\tilde\theta_1 - 3 \tilde\theta_2, \;\;\;\;
\\
C_4(\tilde\theta_1,\tilde\theta_2)= 2 \tilde\theta_2, &
C_5(\tilde\theta_1,\tilde\theta_2)= \tilde\theta_1 - \tilde\theta_2, &
C_6(\tilde\theta_1,\tilde\theta_2)= \tilde\theta_1 + \tilde\theta_2.
\end{array}
\end{equation}
For the trivial background field, $(\tilde\theta_1,\tilde\theta_2)=(0,0)$, 
we have $W= - \frac{14}{45} \pi^2/\beta^4$, which corresponds to an ideal gas 
of $G(2)$ gluons. Here, the factor 14 results from the number of $G(2)$
generators. We like to point out that, since $SU(3)$ is a subgroup of $G(2)$ with
the same rank, the high-temperature effective potential in $SU(3)$ Yang-Mills 
theory \cite{Wei81,Wei82,Bel90} can be immediately obtained from 
eq.(\ref{effpot}) performing the sum only up to $i=3$. The figures
\ref{effpotG2SU3}a and \ref{effpotG2SU3}b show the contour plots of the 
one-loop effective potential for the Polyakov loop at high temperature in 
$SU(3)$ and $G(2)$ Yang-Mills theory, respectively.

Somewhat unexpectedly, the $G(2)$ effective potential still shows traces of the
$\Z(3)$ center of the $SU(3)$ subgroup. While we have an absolute minimum
corresponding to the trivial center element 
$(\tilde\theta_1,\tilde\theta_2)=(0,0)$, we also have local minima 
corresponding to the nontrivial $\Z(3)$ center elements
$(\tilde\theta_1,\tilde\theta_2)=(0,\pm 2\pi/3)$ and 
$(\tilde\theta_1,\tilde\theta_2)=(\pm \pi,\pm \pi/3)$. The nontrivial $\Z(3)$ 
center elements correspond to metastable minima since they are separated from 
the absolute minimum by a free energy gap that grows with the volume. Hence,
the local minima are irrelevant in the thermodynamic limit. However, as
mentioned in the previous section, numerical simulations on finite lattices can
be affected by tunneling to one of these metastable minima. Depending on the 
efficiency of the Monte Carlo algorithm, the metastable phases may have long 
lifetimes and may significantly bias the numerical results.

\section{Deconfinement in the $G(2)$ Gauge-Higgs Model}

Using the Higgs mechanism to interpolate between $SU(3)$ and $G(2)$ Yang-Mills
theory, we will address the issue of the deconfinement phase transition. In 
the $SU(3)$ theory this transition is weakly first order. As the mass of the 6 
additional $G(2)$ gluons is decreased, the $\Z(3)$ center symmetry of $SU(3)$
is explicitly broken and the phase transition is weakened. Qualitatively, we 
expect the heavy gluons to play a similar role as heavy quarks in $SU(3)$ QCD \cite{Has83}. 
Hence, we expect the first order deconfinement phase transition line to
terminate at a critical endpoint before the additional $G(2)$ gluons have 
become massless. As we further lower the mass of the additional 6 
$G(2)$ gluons by decreasing the value of the hopping parameter $\kappa$, a line
of first order phase transitions (connected to the deconfinement phase 
transition of $G(2)$ Yang-Mills theory) reemerges. In contrast to the $SU(3)$
transition, which is an unavoidable consequence of the nontrivial $\Z(3)$
center symmetry, we attribute the $G(2)$ transition to the large mismatch 
between the relevant degrees of freedom between the low- and the 
high-temperature regions. As 6 of the 14 $G(2)$ gluons are progressively 
removed from the dynamics by the Higgs mechanism, this mismatch becomes smaller
and the reason for the phase transition disappears. 

In the $1/g^2 = \infty$ limit the gauge degrees of freedom are frozen and the
lattice theory reduces to the 4-d $SO(7)$ nonlinear $\sigma$-model. This model
has a second order phase transition that separates the symmetric phase at small
values of $\kappa$ from the $SO(7) \rightarrow SO(6)$ broken phase at large 
$\kappa$. At large but finite values of $1/g^2$ this transition is expected to
extend to a line of first order transitions. 

\begin{figure}[htb]
\begin{center}
\vskip-2.mm \hskip.05cm  (a) \hskip4.cm (b) \hskip4.cm (c)\\
\epsfig{file=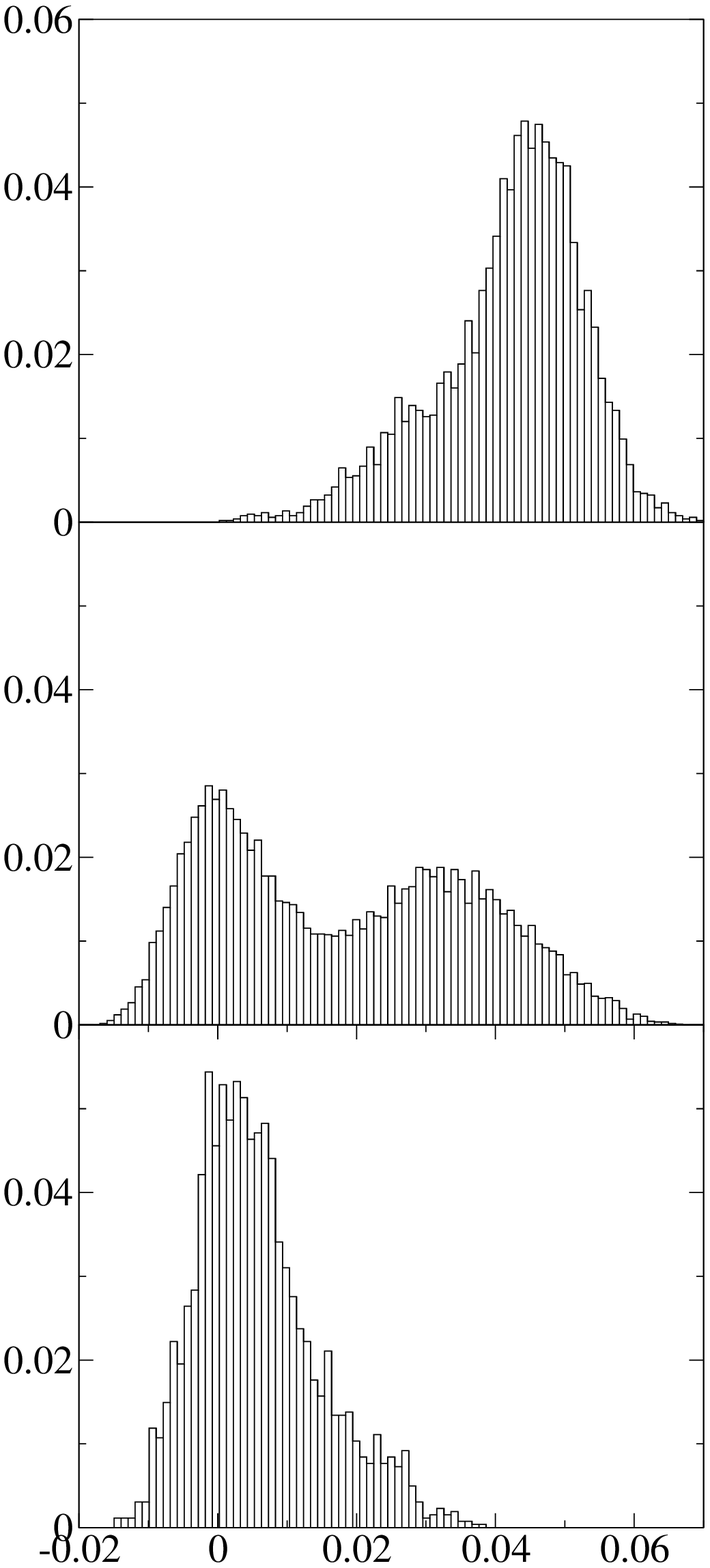,width=3.5cm,height=7.5cm}\hskip1cm
\epsfig{file=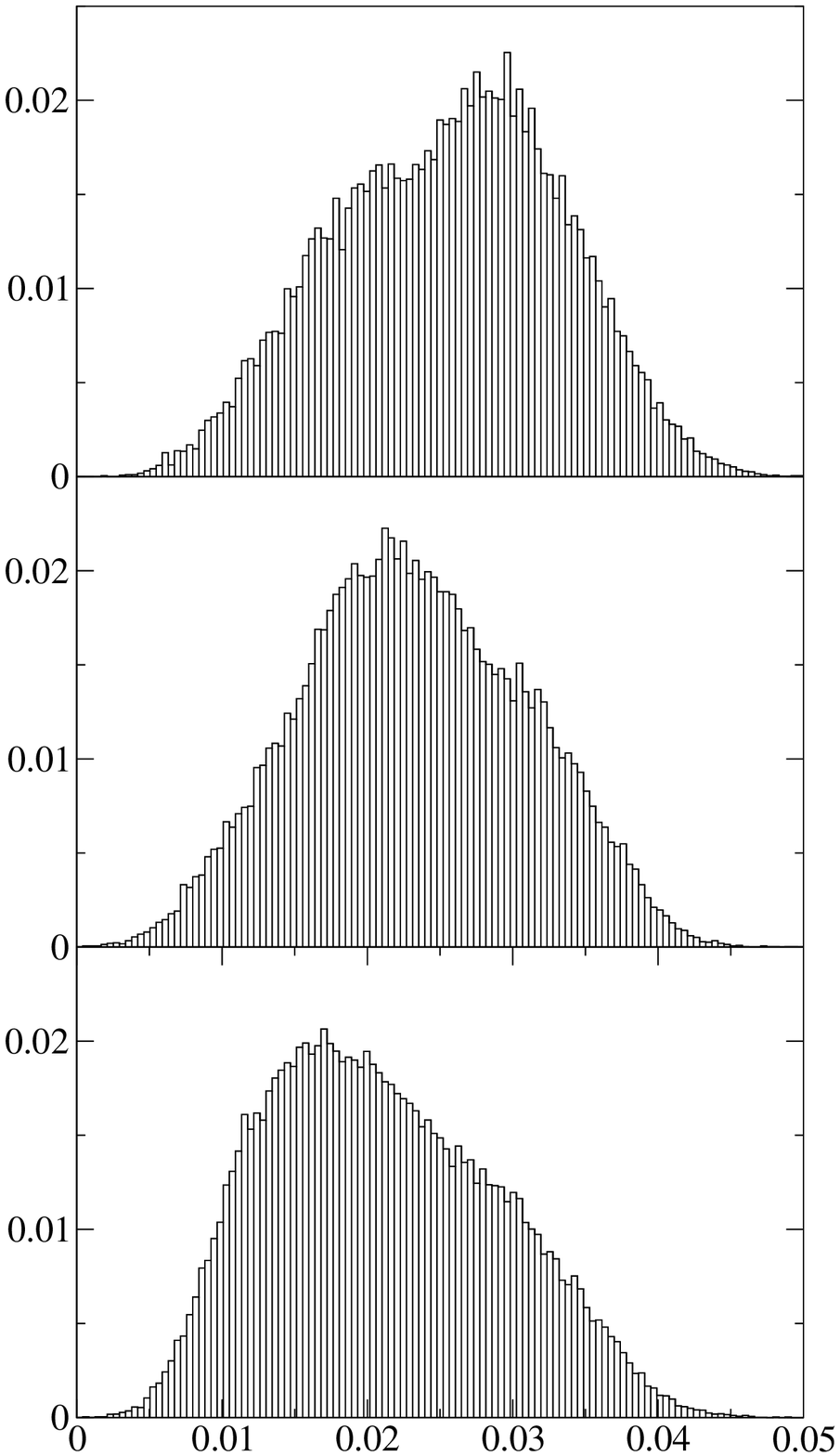,width=3.5cm,height=7.5cm}\hskip1cm
\epsfig{file=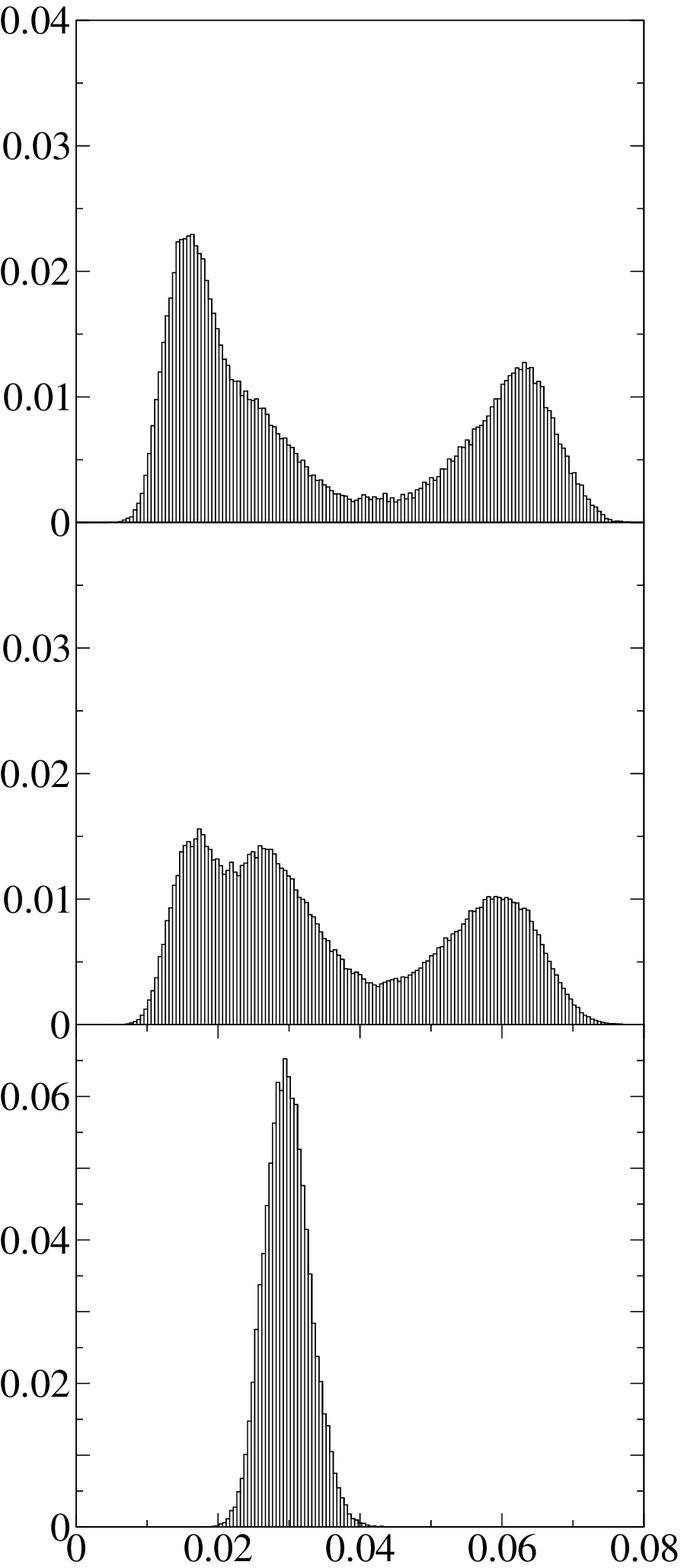,width=3.5cm,height=7.5cm}
\caption{Polyakov loop probability distributions at $\kappa=1.3$ (a), $\kappa=1.5$ (b) and
  $\kappa=4.0$ (c). The temperature increases from bottom to top. The numerical
  simulations at $\kappa=1.3$ have been carried out on a $14^3\times 6$ lattice at the
  gauge couplings $7/g^2=9.7$, 9.72 and 9.73 (bottom to top). The simulations at
  $\kappa=1.5$ and 4.0 have been performed on a $24^3\times 6$ lattice. The gauge couplings are 
 $7/g^2=9.553$, 9.5535 and 9.554~(b) and $7/g^2=8.65$, 8.72 and 8.727~(c) (bottom to
  top).}\label{critG2HYM}
\end{center}
\end{figure}

We have performed numerical simulations on lattices with time extent $N_t = 6$
and spatial sizes $N_s$ ranging from 14 to 24, measuring the probability
distribution of the Polyakov loop. As discussed earlier, figure \ref{YMhist} shows this 
distribution for the $G(2)$ Yang-Mills theory in the region around the phase
transition. The corresponding distributions in the $G(2)$ gauge-Higgs model are
shown in figure \ref{critG2HYM}. 
The three parts of Figure \ref{critG2HYM}a correspond to the same small value of $\kappa = 1.3$ 
at different temperatures (i.e.\ different values of $1/g^2$). As in $G(2)$
Yang-Mills theory, the Polyakov loop is very small in the low-temperature confined region
(bottom panel in figure \ref{critG2HYM}a). As we increase the temperature towards $T_c$, a
deconfined peak with a large value of the Polyakov loop emerges (middle panel in figure
\ref{critG2HYM}a), and finally at high temperatures the confined peak disappears (top
panel in figure \ref{critG2HYM}a). At larger values of the hopping parameter ($\kappa =1.5$) 
the situation is different. As displayed in figure \ref{critG2HYM}b there is now
only one peak (although there is a small bump in its shoulder). As the temperature is
increased, the peak progressively broadens and then, in a restricted range of intermediate
temperatures, it changes its skewness from positive (bottom panel in figure
\ref{critG2HYM}b) to negative (top panel in figure \ref{critG2HYM}b). When the temperature
is increased further, the peak shrinks again in the deconfined region. At even larger
values of the hopping parameter ($\kappa = 4.0$) the situation changes again. The 6
additional $G(2)$ gluons are now quite heavy and begin to decouple from the dynamics of
the remaining 8 $SU(3)$ gluons. As a result, the $\Z(3)$ center emerges as an approximate
symmetry which becomes exact in the pure $SU(3)$ limit. This manifests itself by an
additional peak due to the presence of metastable $\Z(3)$ images of the deconfined phase
(top panel in figure \ref{critG2HYM}c). The metastable states become stable phases only at
$\kappa = \infty$. It should be noted that $SU(3)$ is embedded as $\{3\} \oplus
\{\overline{3}\} \oplus \{1\}$ in the real 7-dimensional representation of $G(2)$. The
additional peak thus represents the sum of the two nontrivial $\Z(3)$ images of the
deconfined phase which explains its large height. Still, at finite $\kappa$, the presence
of the additional metastable peak is a finite size effect that disappears in the infinite
volume limit. Note that, for very large values of $\kappa$, the critical coupling of the
$G(2)$ Gauge-Higgs model at $N_t=6$ has to approach the critical coupling of the $SU(3)$
Yang-Mills theory at the same $N_t$. Taking into account a factor 7/6 due to the different
normalization of the trace between $G(2)$ and $SU(3)$, we have indeed verified this
feature.

\section{Conclusions}

We have investigated the deconfinement phase transition in $G(2)$ gauge 
theory. This is interesting because $G(2)$ is the smallest simply connected 
group with a trivial center. Due to the trivial center, there is no symmetry
reason for the existence of a finite temperature phase transition in $G(2)$
Yang-Mills theory. Still, using Monte Carlo simulations, we have found a clear
signal for a first order deconfinement phase transition. Interestingly, $SU(3)$
is a subgroup of $G(2)$. Hence, exploiting the Higgs mechanism, $G(2)$ can be
broken down to $SU(3)$, and one can thus interpolate between $G(2)$ and $SU(3)$
Yang-Mills theories. Remarkably, the $G(2)$ phase transition disappears (or at
least weakens substantially) before the $SU(3)$ transition emerges. From this
observation we conclude that the $G(2)$ and $SU(3)$ deconfinement phase 
transitions arise for two different reasons. The $SU(3)$ transition is an
unavoidable consequence of the $\Z(3)$ center symmetry which gets spontaneously
broken in the high-temperature phase. As the 6 additional $G(2)$ gluons 
(transforming as $\{3\}$ and $\{\overline{3}\}$ of $SU(3)$) enter the dynamics,
the $\Z(3)$ center symmetry is explicitly broken. Just as in QCD with heavy
quarks, this leads to a weakening of the deconfinement phase transition and
possibly to its termination in a critical endpoint. We attribute the existence
of the $G(2)$ transition to a different mechanism: the transition arises due to
a large mismatch in the number of the relevant degrees of freedom at low
and high temperatures. In particular, in the confined phase there is a small
number of glueball states, while in the deconfined phase there is a large
number of 14 deconfined gluons. When the Higgs mechanism gives mass to 6 of 
these gluons, they are progressively removed from the dynamics. Consequently,
the mismatch in the number of degrees of freedom is reduced and the reason for
the $G(2)$ deconfinement phase transition disappears.

Our study contributes to the investigation of the deconfinement phase transition in
Yang-Mills theories with a general gauge group. Remarkably, in $(3+1)$ dimensions only
$SU(2)$ Yang-Mills theory has a second order phase transition. In particular, $SU(N)$
Yang-Mills theories with any higher $N$ seem to have first order deconfinement phase
transitions. Although all $Sp(N)$ Yang-Mills theories have the same nontrivial center
$\Z(2)$, only $Sp(1) = SU(2)$ has a second order phase transition, while $Sp(2)$, $Sp(3)$,
and most likely all higher $Sp(N)$ Yang-Mills theories have a first order transition. As
for $G(2)$, we have attributed this to the large number of deconfined gluons in the
high-temperature phase \cite{Hol03a,Hol03b,Pep04}. Let us also discuss $SO(N)$ (or more
precisely $Spin(N)$) Yang-Mills theories. Since $SO(3) \simeq Spin(3) = SU(2)$, 
$SO(4) \simeq Spin(4) = SU(2) \otimes SU(2)$, $SO(5) \simeq Spin(5) = Sp(2)$, and 
$SO(6) \simeq Spin(6) = SU(4)$, we can limit the discussion to $SO(7)$ and higher. As far
as we know, such theories have never been simulated. Still, due to the large size of the
gauge group (just like $Sp(3)$, $SO(7)$ has 21 generators), we expect a strong first order
deconfinement phase transition.  This finally leads us to the other exceptional groups
$F(4)$, $E(6)$, $E(7)$, and $E(8)$. Interestingly, two of these --- namely $F(4)$ and
$E(8)$ --- have a trivial center, just like $G(2)$, while $E(6)$ and $E(7)$ have
nontrivial centers $\Z(3)$ and $\Z(2)$, respectively. However, independent of the center,
since all these groups have a large number of generators, we again expect strong first
order deconfinement phase transitions.

\section*{Acknowledgements}

We dedicate this work to Peter Minkowski on the occasion of his $65^{th}$ birthday.
He was first to point out to one of us that $G(2)$ has a trivial center. This 
paper is based on previous work with him on exceptional confinement in $G(2)$ 
gauge theory. We like to thank P.~de~Forcrand, K.~Holland, O.~Jahn, and 
P.~Minkowski for interesting discussions. This work is supported in part by the 
Schweizerischer Nationalfonds.

\begin{appendix}

\section{Heat Bath Algorithm for $G(2)$ Lattice Gauge Theory}

There are several ways to simulate $G(2)$ lattice gauge theory. In particular,
Cabbibo and Marinari's method of updating embedded $SU(2)$ subgroups suggests
itself. There are two ways in which $SU(2)$ subgroups can be embedded in  $G(2)$. In the
first case, the fundamental representation of $G(2)$ decomposes as 
\begin{equation}
\{7\} = 2 \{2\} \oplus 3 \{1\}.
\end{equation}
This applies to those three $SU(2)$ subgroups of $G(2)$ that are 
simultaneously subgroups of the $SU(3)$ embedded in $G(2)$. Then, just as in 
$SU(3)$, Cabbibo and Marinari's method can be applied using a heat bath 
algorithm. In this way we have updated the $SU(3)$ subgroup of $G(2)$. 
However, there is a second case in which the fundamental representation of 
$G(2)$ decomposes as
\begin{equation}
\{7\} = 2 \{2\} \oplus \{3\}.
\end{equation}
Due to the presence of the $SU(2)$ adjoint representation $\{3\}$, in this case
one cannot apply the heat bath algorithm. Instead of applying the Metropolis
algorithm, we use a look-up table of randomly distributed $G(2)$ gauge
transformations (together with their inverses) to rotate the $SU(3)$ subgroup
through $G(2)$ \footnote{We thank P.\ de Forcrand for suggesting this 
procedure.}. Combined with the heat bath update of the $SU(3)$ subgroup 
described above, this method is ergodic and obeys detailed balance. 
Alternatively, one could use just a Metropolis algorithm or perform updates in
$U(1)$ subgroups.

Due to round-off errors, it is important to occasionally project the link variables back
into $G(2)$. This can be accomplished in a straightforward manner by going to the algebra 
of $G(2)$. However, there is a more efficient method that uses an explicit representation of
the $G(2)$ matrices~\cite{Mac02} based on the fact that $SU(3)$ is a subgroup of $G(2)$. In
fact, a $G(2)$ matrix $\Omega$ can be expressed as the product of two matrices 
$\Omega = {\cal {Z}}\; {\cal{U}} $.  The matrix ${\cal{U}}$ belongs to the $SU(3)$
subgroup while the matrix ${\cal {Z}}$ is in the coset $G(2)/SU(3) \sim S^6$ describing
a 6-dimensional sphere. These matrices are given by
\begin{equation}
{\cal {Z}}(K) = \left(
\begin{array}{ccc} C(K) & \mu(K)\, K & D(K)^* \\
-\mu(K)\, K^\dagger & \frac{1-|K|^2}{1+|K|^2} & -\mu(K)\, K^T\\
D(K) & \mu(K)\, K^* & C(K)^*
\end{array} \right),
\hskip .5cm
{\cal {U}}(U) = \left(
\begin{array}{ccc} U & 0 & 0 \\
0 &  1 & 0\\ 0 & 0 &\; U^*
\end{array} \right),
\end{equation}
where $K$ is a 3-component complex vector and $U$ is an $SU(3)$ matrix. The
number $\mu(K)$ is given by $\mu(K)=\sqrt{2}/(1+|K|^2)$,
while the $3 \times  3$ matrices $C(K)$ and $D(K)$ are
\begin{equation}
C(K)= \frac{1}{\Delta}\left\{ \1 -\frac{K\,K^\dagger}{\Delta (1+\Delta)} \right\},
\hskip 1.2cm
D(K) = -\frac{W}{\Delta}-\frac{K^*\,K^\dagger}{\Delta^2},
\end{equation}
where
\begin{equation}
W_{\alpha\beta}= \epsilon_{\alpha\beta\gamma} K_\gamma, 
\;\;\;\;\;\;
\Delta=\sqrt{1+|K|^2}.
\end{equation}

The projection can be performed by obtaining the number $\mu$ and the
vector $K$ from the matrix element $\Omega_{44}$ and the 3-component vector
$(\Omega_{14},\Omega_{24},\Omega_{34})$. The matrix $U$ can then be derived from the
$3\times 3$ matrices $\Omega_{ij}$ with $i,j=1,2,3$ and 
$C^{-1}=\Delta (\1 + K\,K^\dagger/(1+\Delta))$.
Finally, we note that using this parametrization one can see that $\Omega$ indeed depends
on 14 real parameters since the vector $K$ and the $SU(3)$ matrix $U$ depend,
respectively, on 6 and 8 real parameters. 

\end{appendix}

\end{document}